\documentclass[floatfix,twocolumn,nofootinbib,superscriptaddress,a4paper,preprintnumbers]{revtex4-1}

\usepackage{amsmath}
\usepackage{amssymb}
\usepackage{graphicx}
\usepackage{ulem}
\usepackage[mathscr]{eucal}
\usepackage{color}
\usepackage[T1]{fontenc} 
\usepackage{slashed}
\usepackage{xspace}
\usepackage{hyperref}
\usepackage{xspace}
\usepackage{listings}
\usepackage{gensymb}

\newcommand\Vevacious{{\tt Vevacious}\xspace}

\newcommand\SARAH{{\tt SARAH}\xspace}
\newcommand\SPheno{{\tt SPheno}\xspace}

\newcommand{\DRbar}{$\overline{\text{DR}}'$}

\begin{document}

\preprint{KA-TP-33-2018}

\title{Theoretical Constraints on Supersymmetric Models: \\ Perturbative Unitarity vs. Vacuum Stability}

%
\author{Florian Staub}
\email{florian.staub@kit.edu}
\affiliation{Institute for Theoretical Physics (ITP), Karlsruhe Institute of Technology, Engesserstra{\ss}e 7, D-76128 Karlsruhe, Germany}
\affiliation{Institute for Nuclear Physics (IKP), Karlsruhe Institute of Technology, Hermann-von-Helmholtz-Platz 1, D-76344 Eggenstein-Leopoldshafen, Germany}

\begin{abstract}
There are nowadays strong experimental constraints on supersymmetric theories from the 
Higgs measurements as well as from the null results in Sparticle searches. However, even the parameter spaces which are in agreement with experimental data can be
further constrained by using theoretical considerations. Here, we discuss for the MSSM and NMSSM the impact of perturbative unitarity as well as of the stability 
of the one-loop effective potential. We find in the case of the MSSM, that vacuum stability is always the stronger constraint. On the other side, the situation is more diverse 
in the NMSSM and one should always check both kind of constraints. 
\end{abstract}
\maketitle

\section{Introduction}
The discovery of a standard model (SM)-like Higgs boson with a mass of about 125~GeV \cite{Aad:2012tfa,Chatrchyan:2012xdj} seems to be a good argument that supersymmetry (SUSY), and in particular the minimal supersymmetric standard model (MSSM), is the correct extension of the SM:  in contrast to other ideas to extend the SM, the MSSM predicts that the Higgs boson shouldn't be significantly heavier than the $Z$-boson if new physics comes into play at the TeV scale, see e.g. Ref.~\cite{Martin:1997ns} and references therein. Such a mass scale for SUSY particles would also solve the hierarchy problem of the SM to a large extent. This is a big advantage compared to other ideas of beyond-the-SM (BSM) physics like technicolor. In technicolor the natural mass range for the Higgs lies at scales well above the measured mass. However, a closer look  shows that the situation is more complicated in the MSSM. The Higgs mass requires large radiative corrections in order to be compatible with experimental data. The main source of these corrections are the superpartners of the top, the stops. In order to maximise their contributions to the Higgs mass, one needs to consider scenarios in which they are maximally mixed \cite{Brignole:1992uf,Chankowski:1991md,Dabelstein:1994hb,Pierce:1996zz}. This can be dangerous because of two reasons. The first and better studied issue is that a large stop mixing can cause the presence of charge- and colour-breaking (CCB) minima in the scalar potential \cite{Camargo-Molina:2013sta,Blinov:2013fta,Chowdhury:2013dka,Camargo-Molina:2014pwa}. Since the tunnelling rate to these vacua is typically large, this results in tension between an acceptable Higgs mass and a sufficiently long-lived electroweak (ew) breaking vacuum. The second problem with highly mixed stops is that perturbative unitarity could be violated: the large trilinear stop couplings which are responsible for the mixing can induce scalar scattering processes which violate unitarity at leading order. While this is cured at higher loop-level, it indicates a breakdown of perturbation theory. The impact of these constraints on trilinear couplings in the MSSM was only discussed in a conference note up to now \cite{Schuessler:2007av}. \\
One can alleviate the need for large loop corrections from the stop sector by considering SUSY models in which the Higgs mass is already enhanced at tree-level. The simplest extension in this direction is to add a gauge singlet, resulting in the next-to-minimal supersymmetric standard model (NMSSM). The additional $F$-term contributions in the NMSSM can raise significantly the tree-level Higgs mass \cite{Ellwanger:2009dp,Ellwanger:2006rm}. Therefore, the vacuum stability problems of the MSSM due to CCB minima are cured. However, the extended Higgs sector in the NMSSM introduces new couplings which can potentially destabilise the ew vacuum. The global minimum of the scalar potential might still be charge conserving, but the mass of the $Z$-boson would be completely different. Large trilinear couplings or light states in the extended Higgs sector can also cause scattering cross sections which violate perturbative unitarity at leading order. The vacuum stability in the NMSSM has been studied in the past at tree-level \cite{Ellwanger:1996gw, Ellwanger:1999bv, Kanehata:2011ei, Kobayashi:2012xv,Agashe:2012zq}, and also with one-loop corrections \cite{Beuria:2016cdk,Krauss:2017nlh,Beuria:2017wox} in quite some detail. In contrast, the limits from perturbative unitarity are again less explored also in the NMSSM \cite{Betre:2014fva} and hardly included in phenomenological studies.\\
The aim of this letter is to discuss the importance of perturbative unitarity and vacuum stability in the MSSM and NMSSM. It will be shown that a proper check of the vacuum stability in the MSSM might be sufficient to identify all dangerous parameter regions. On the other side, it turns out that in the NMSSM one should always check both constraints. We start in Sec.~\ref{sec:constraints} with a brief discussion of the two constraints in general. In Sec.~\ref{sec:results} the main results for the MSSM and NMSSM are summarised. We conclude in Sec.~\ref{sec:conclusion}.

\section{Unitarity and vacuum stability}
\label{sec:constraints}
\subsection{Calculation of the unitarity constraints}
Perturbative unitarity gives constraints on the maximal size of  $2 \to 2$ scalar field scattering amplitudes. 
 The 0th partial wave amplitude $a_0$ is given by
\begin{equation}
a_0^{ba} = \frac{1}{32\pi} \sqrt{\frac{4 |\vec{p}^{\,b}| |\vec{p}^{\,a}|}{2^{\delta_{12}} 2^{\delta_{34}}\, s}} \int_{-1}^1 d(\cos \theta) \mathcal{M}_{ba} (\cos \theta)\,,
\end{equation}
where $\vec p^{\,a(b)}$ is the center-of-mass three-momentum of the incoming (outgoing) particle pair $a=\{1,2\}~(b=\{3,4\})$, $\theta$ is the angle between these three-momenta and $\mathcal M_{ba}(\cos\theta)$ is the scattering matrix element. $a_0$ must satisfy 
\begin{equation}
|a_0| \leq 1 \hspace{1cm} \text{or} \hspace{1cm} |{\mathcal Re} [ a_0]| \leq \frac{1}{2}
\end{equation}
The amplitudes are real at tree-level. This gives the more severe constraint  $|{\mathcal Re} [ a_0]| \leq \frac{1}{2}$. This usually gets simplified  in the limit 
$s\to \infty$ to $|\mathcal{M}| < 8 \pi$. This condition must be  satisfied by any eigenvalue $\tilde{x}_i$ of the scattering matrix $\mathcal{M}$, where $\mathcal{M}$ is derived by  including all possible combinations of scalars in the initial and final state, i.e. the scattering matrix in BSM models is usually quite large. Therefore, often approximations applied to obtain analytical results. One widely used ansatz is to consider the limit $s\to \infty$. However, it has recently been pointed out that this approximation is often not satisfied, see Refs.~\cite{Goodsell:2018tti,Goodsell:2018fex,Krauss:2018orw}. Instead, one should scan over $s$ and check for the maximal amplitude $a_0$ at given $s$. This is also the only possible to keep the impact of trilinear scalar couplings which are crucial for checking perturbative unitarity in the MSSM and NMSSM. In contrast to the large $s$ approximation, this approach introduces the additional difficulty that poles can appear. In this work, the poles are handled carefully as discussed in Refs.~\cite{Schuessler:2007av,Goodsell:2018tti} to obtain amplitudes which are not artificially enhanced. 

\subsection{Checking the vacuum stability}
In order to check the stability of the electroweak vacuum, we consider the one-loop effective potential given by
\begin{equation}
V_{EP}^{(1)} = V_{\rm Tree} + V^{(1)}_{\rm CT}  + V^{(1)}_{\rm CW} 
\end{equation}
For a stable ew vacuum, the deepest minimum in this potential must coincide with the ew one. Here, $V_{\rm Tree}$ is the tree-level potential which for supersymmetric theories consists of
\begin{equation}
V_{\rm Tree} = V_F + V_D + V_{\rm soft} 
\end{equation}
While the $F$-and $D$-term potentials are by construction positive, the soft-breaking potential $V_{\rm soft}$ can destabilise the vacuum. For a long time, 
the vacuum stability in the MSSM was checked only taking $V_{\rm Tree}$ into account and considering just $D$-flat directions, i.e. $V_D \simeq 0$. However, it has been 
pointed out in Ref.~\cite{Camargo-Molina:2013sta} that loop corrections can be important and that the global minimum of the potential must not be located in a $D$-flat direction. Therefore, we are 
going to include also the other two terms which are necessary to encode the loop corrections and we are going to search numerically for the global minimum of the potential. 
$V^{(1)}_{\rm CT}$ is the counter-term (CT) potential. Since the calculation of the SUSY and Higgs masses is performed in the \DRbar scheme, also the CTs must be calculated in this scheme. CTs appear only for the parameters  which are obtained from the tadpole conditions. In our case, these are the soft-breaking terms $m_i^2$ in the Higgs sector. The renormalisation conditions at one-loop level are
\begin{equation}
\delta t_i + t_i(m_i^2 \to m_i^2 + \delta m_i^2)   \equiv 0\\ 
\end{equation}
where $t_i$ are the tree-level tadpole conditions $t_i = \frac{\partial V}{\partial \phi_i}$ and $\delta t_i$ are the one-loop corrections. Finally, the Coleman-Weinberg 
potential $V^{(1)}_{\rm CW}$ is given by \cite{Coleman:1973jx}
\begin{equation}
V^{(1)}_{\rm CW} = \frac{1}{16\pi^2} \sum_{i}^{\text{all fields}} r_i s_i C_i m_i^4 \left(\log\frac{m_i^2}{Q^2} - \frac32 \right)
\end{equation}
with $r_i = 1$ for real bosons, otherwise 2; $C_i =3$ for quark, otherwise 1; $s_i=-\frac{1}{2}$ for fermions, $\frac{1}{4}$ for scalars 
and $\frac{3}{4}$ for vector bosons. More details about the calculation and the numerical approach to find the global minimum are given in Refs.~\cite{Camargo-Molina:2013sta}.\\
Even if the global minimum is not the ew one, this doesn't immediately rule out a parameter point. It could still be, that the ew vacuum is long lived on cosmological time scales. Therefore, it's necessary in these cases to calculate the tunnelling rate. We are going to use the approach discussed in Ref.~\cite{Camargo-Molina:2014pwa}, but we don't include thermal effects. The reason is that one can avoid the faster tunnelling due to thermal effects by assuming a reheating temperature after inflation which is sufficiently low. A parameter point is considered as 'long-lived' if its life-time is longer than the age of the universe.

\section{Results}
\label{sec:results}
Our numerical analysis is based on the tool-chain \SARAH--\SPheno--\Vevacious--{\tt CosmoTransitions}. We used \SARAH \cite{Staub:2008uz,Staub:2009bi,Staub:2010jh,Staub:2012pb,Staub:2013tta}  to generate \SPheno modules \cite{Porod:2003um,Porod:2011nf} for the MSSM and NMSSM. With this module we calculate the SUSY and Higgs masses including two-loop corrections \cite{Goodsell:2014bna,Goodsell:2014pla,Goodsell:2015ira,Goodsell:2016udb}. 
\SPheno is also the only available code which calculates the scattering amplitudes at finite $s$  \cite{Goodsell:2018tti}\footnote{The public version of \SARAH/\SPheno include so far only the possibility of colour singlets in the initial and final state. This has been generalised for this project to be able to include all processes involving stops in the checks for perturbative unitarity. The changes will be discussed and published elsewhere. }.
The spectrum file generated by \SPheno is passed as input for \Vevacious \cite{Camargo-Molina:2013qva}. \Vevacious finds all solutions to the tree-level tadpole equations by using a homotopy continuation implemented in the code {\tt HOM4PS2}. These minima are used as the starting points to find the minima of the one-loop effective potential using {\tt minuit} \cite{James:1975dr}. If it finds deeper minima than the ew one, \Vevacious calls {\tt CosmoTransitions} \cite{Wainwright:2011kj} to get the tunnelling rate. 

\subsection{MSSM}
We start with the MSSM, see Refs.~\cite{Nilles:1983ge,Martin:1997ns} and references therein for detailed discussions of this model. We just briefly summarise our conventions.  The most general renormalisable and SM gauge invariant superpotential which also respects $R$-parity reads
\begin{align}
 W_{\rm MSSM}=& Y_u \hat{u} \hat{q} \hat{H}_u - Y_d \hat{d} \hat{q} \hat{H}_d - Y_e \hat{e} \hat{l} \hat{H}_d \nonumber \\
 & + \mu \hat{H}_u \hat{H}_d
\end{align}
where we suppressed flavour and $SU(2)$ indices. $Y_e,\,Y_d,\,Y_u$ are dimensionless
3x3 matrices of Yukawa couplings.  The
soft SUSY breaking potential reads
\begin{eqnarray}
V_{\rm soft}  &=& m_{H_u}^2 |H_u|^2 + m_{H_d}^2
|H_d|^2+ \tilde{f}^\dagger m_{\tilde{f}}^2 \tilde{f} \nonumber \\ && + \frac{1}{2}\left(M_1 \, \tilde{B}
\tilde{B} + M_2 \, \tilde{W}_i \tilde{W}^i + M_3 \, \tilde{g}_\alpha
\tilde{g}^\alpha + h.c.\right) \nonumber \\ && 
+ \Big(T_u \tilde{u} \tilde{q} {H}_u - T_d \tilde{d} \tilde{q} {H}_d - T_e \tilde{e} \tilde{l} {H}_d \nonumber \\
&& \hspace{1cm} + B_\mu {H}_u {H}_d + h.c.\Big)
\end{eqnarray}
with $\tilde{f}=\{\tilde{d},\tilde{u},\tilde{q},\tilde{l},\tilde{e}\}$.
Since flavour effects are negligible for the discussions 
in the following, we make the assumption that all Yukawa and soft-breaking matrices are diagonal. Moreover, only third generation couplings 
are important We can then introduce the common parametrisation
of the trilinear soft-couplings:
\begin{equation}
T_i = A_i Y_i \hspace{1cm} \text{for}\quad i=\{t,b,\tau\}
\end{equation}
After EWSB, the neutral components of the Higgs doublets receive vacuum expectation values (VEVs) $v_d$, $v_u$ with $v=\sqrt{v_d^2+v_u^2} \simeq 246$~GeV and $\tan\beta=\frac{v_u}{v_d}$. 
Among many other things, 
this causes a mixing in the stop sector. The mass matrix of the two stops is given in the basis $(\tilde{t}_L, \tilde{t}_R)$ by
\begin{align}
 m^2_{\tilde t} =&\left( 
\begin{array}{cc}
D + m_t^2 +  m_{\tilde{t}_L}^2  &\frac{1}{\sqrt{2}}  Y_{t} \Big(v_u A_{t}- v_d \mu  \Big) \\ 
\frac{1}{\sqrt{2}}  Y_t \Big( v_u A_t  - v_d \mu  \Big) & m_{\tilde{t}_R}^2 + D' + m_t^2 \end{array} 
\right) 
\end{align}
where the explicit form of $D$-term contributions is skipped for brevity because these terms are usually sub-dominant. \\
Diagonalisation results in two physical mass eigenstates $\tilde{t}_1$ and $\tilde{t}_2$ and one mixing angle $\Theta$. The value of the angle is given by
\begin{align}
\sin2\theta =  -\frac{\sqrt{2} Y_t \Big( v_u A_t  - v_d \mu  \Big)}{m_{t_2}^2-m_{t_1}^2}
\end{align}
At tree-level the light SM-like Higgs mass is bounded in the MSSM by $m_h^2 \leq M_Z^2$ and the additional heavy Higgs states have a mass of
$M_A = \frac{1+\tan\beta^2}{\tan\beta} B_\mu$. The main loop corrections to the light Higgs mass in the MSSM stem from the (s)top contributions. They 
can be written in the decoupling limit $M_A \gg M_Z$  at one-loop-level as 
\cite{Haber:1993an,Carena:1995wu,Martin:1997ns,Heinemeyer:1998np,Heinemeyer:1999be,Carena:2000dp}
\begin{equation}
\delta m_h^2 =\frac{3}{2 \pi^2} \frac{m_t^4}{v^2 }\left[\log \frac{M_{\rm SUSY}^2}{m_t^2} + \frac{X_t^2}{M_{\rm SUSY}^2} \left(1 - 
\frac{X_t^2}{12 M_{\rm SUSY}^2}\right) \right] 
\label{eq:higgs-stop}
\end{equation}
with $M_{\rm SUSY} \equiv \sqrt{m_{\tilde{t}_1} m_{\tilde{t}_2}}$,  $m_t$ being the running \DRbar top mass and 
$X_t \equiv A_t - \mu\cot\beta$. One can see from eq.~(\ref{eq:higgs-stop}) that the one-loop corrections are maximised for 
$|X_t| = \sqrt{6} M_{\rm SUSY}$, while they quickly drop for $|X_t| \gg \sqrt{6} M_{\rm SUSY}$. For $\tan\beta \gg 1$, the contribution from $A_t$ dominates, i.e. 
large trilinear stop couplings are needed to rise the Higgs mass. 

\subsubsection{Stop Scattering}
\begin{figure}
\includegraphics[width=0.7\linewidth]{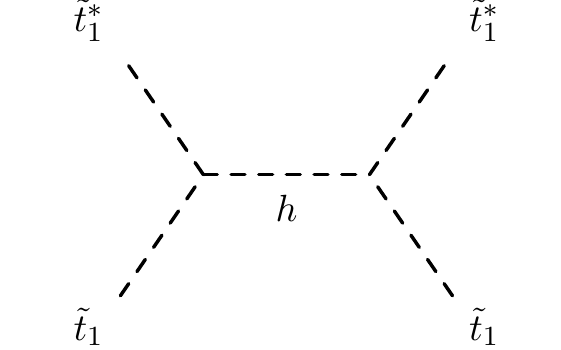} 
\caption{Diagram contributing to stop scattering via Higgs exchange.}
\label{fig:t1t1_scatter}
\end{figure}

\begin{figure}
\includegraphics[width=\linewidth]{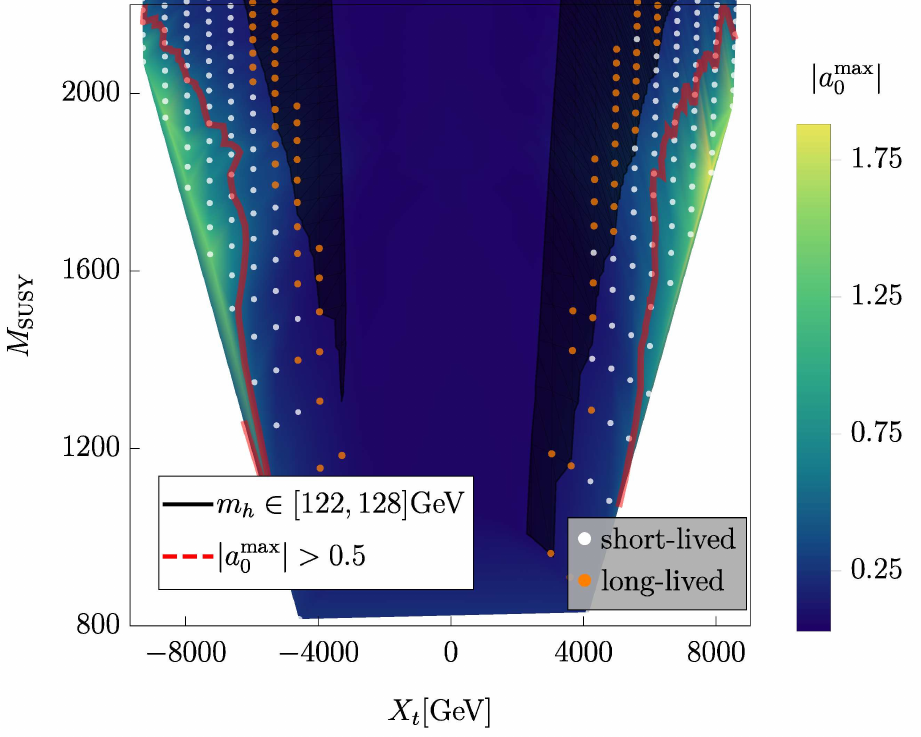} 
\caption{The values of $|a^{\rm max}_0|$ in the ($X_t$,$M_{\rm SUSY})$ plane. The red line indicates $|a^{\rm max}_0|>\frac12$ The black shaded region shows the region with $m_h =125\pm 3$~GeV. The white (orange)  dots refer to an unstable ew vacuum which is short-lived (long-lived) on cosmological time scales.}
\label{fig:MSSM_stop}
\end{figure}

As we have just seen,  large loop corrections to the SM-like Higgs mass need large values for $|A_t|$. These couplings will induce large scattering
cross sections for processes like $\tilde{t}_1 \tilde{t}_1^* \to \tilde{t}_1 \tilde{t}_1^*$ via diagrams shown in Fig.~\ref{fig:t1t1_scatter}.
The amplitude of this process is given by at leading order by
\begin{align}
& a_0^{\tilde{t}_1\tilde{t}_1^*\to\tilde{t}_1\tilde{t}_1^*} = -\sin ^2(2 \Theta_{\tilde{t}})  Y_t (\mu  Y_t \sin\alpha+A_t \cos\alpha)^2 \nonumber \\
& \times \frac{ \left(\left(m_h^2-s\right) \log \left(\frac{m_h^2}{m_h^2-4 m_{\tilde{t}_1}^2+s}\right)+12 m_{\tilde{t}_1}^2-3 s\right)}{32 \pi 
   \left(s-m_h^2\right) \sqrt{s \left(s-4 m_{\tilde{t}_1}^2\right)}}
\end{align}
with the mixing angle $\alpha$ for the CP even Higgs scalars. This can be further simplified to
\begin{align}
a_0^{\tilde{t}_1\tilde{t}_1^*\to\tilde{t}_1\tilde{t}_1^*} 
& \simeq \frac{v^2 X_t^4 Y_t^4 \left(s \log \left(\frac{m_h^2}{m_h^2-4 m_{\tilde{t}_1}^2+s}\right)+12 m_{\tilde{t}_1}^2-3 s\right)}{16 \pi  s
   \left(m_{\tilde{t}_1}^2-m_{\tilde{t}_2}^2\right)^2 \sqrt{s \left(s-4 m_{\tilde{t}_1}^2\right)}}  
\end{align}
In order to get some feeling for the necessary size of $X_t$ which would be in conflict with perturbative unitarity, we assume for the moment $m_{\tilde{t}_1} = 2 m_h$ and $s$ close to the kinematic threshold. That leads to 
\begin{equation}
\frac{X_t}{M_{\rm SUSY}} \Big|_{a_0=\frac12} \simeq 12 \frac{M_{\rm SUSY}}{\text{TeV}} 
\end{equation}
i.e. this ratio is much bigger than $\sqrt{12}$ for reasonable values of $M_{\rm SUSY}$. For these values of $X_t$ we are far away from the preferred window to explain the mass. In Fig.~\ref{fig:MSSM_stop} we check this estimate against a full numerical calculation. Here, the two-loop corrections to $m_h$ as well as all possible scattering processes are included. 
%
We scanned over $m_{\tilde{t}_R}=[0.2,2]$~TeV, and $A_t=[-7,7]$~TeV. $\sqrt{s}$ is always varied between 250~GeV and 5~TeV. The other parameters have been fixed to 
\begin{eqnarray}
&\tan\beta=10,\ M^2_A=2.5~\text{TeV}^2,\ \mu=0.5~\text{TeV}& \nonumber \\
& m_{\tilde{t}_L} =1.8~\text{TeV}&  \nonumber \\
& M_1=0.2~\text{TeV},\ M_2=0.5~\text{TeV},\ M_3=2~\text{TeV}& 
\label{eq:default_values}
\end{eqnarray}
Here and in the following all other sfermion soft-breaking masses are set to 2~TeV and $A_b=A_\tau=0$ is always used if not noted otherwise. \\
We see that perturbative unitarity is only violated in regions where the Higgs mass is much too small. In the regions with $m_h \simeq 125$~GeV the size of $|a_0|$ is at most 0.35. We can now compare this result with the constraints stemming from the stability of the ew potential. This is also depicted in Fig.~\ref{fig:MSSM_stop}. Here, we find that the vacuum stability constraints cut into the interesting regions: some parts of the parameter space with the desirable Higgs mass of $125\pm 3$~GeV suffer from an unstable vacuum. The tunnelling rate at $T=0$ is usually quite small in these regions, i.e. the lifetime of the ew vacuum exceeds the age of the universe and the points could still be assumed to be viable. However, one has to keep in mind that this assumes a reheating temperature after inflation much lower than $M_{\rm SUSY}$. Otherwise, thermal corrections can cause a much larger tunnelling rate \cite{Camargo-Molina:2014pwa}.

\subsubsection{Stau Scattering}
\begin{figure}
\includegraphics[width=\linewidth]{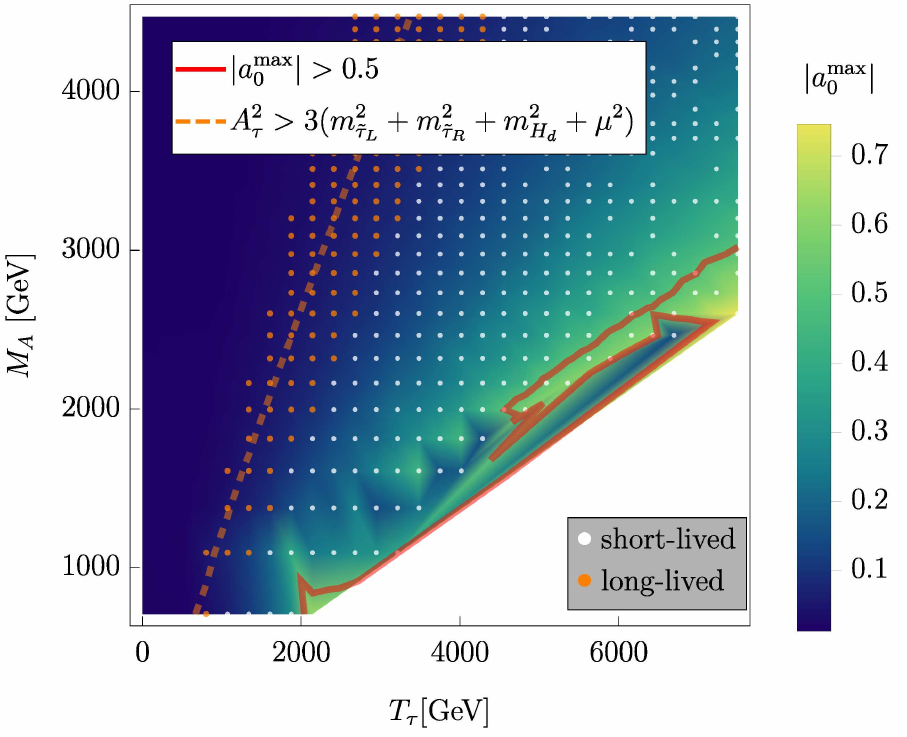} 
\caption{Similar to Fig.~\ref{fig:MSSM_stop} for the ($M_A$, $A_\tau$) plane. We show here also the constraints from vacuum stability based on the approximation eq.~(\ref{eq:thumb_stau}) (orange dashed line). The Higgs mass fulfills in the entire plane $m_h = 125\pm 3$~GeV.}
\label{fig:MSSM_stau}
\end{figure}

Large trilinear couplings can also be present in the slepton sector of the MSSM. This results in light staus which can have interesting phenomenological consequences like forming a co-annihilation region with the neutralino dark matter candidate \cite{Ellis:1999mm}. The coupling $A_\tau$ responsible for the stau mixing can play a similar role as $A_t$ in the stop sector and lead to large scattering rates via the exchange of the heavy Higgs states with mass $M_A$. 
Since the impact of the stau sector on the SM-like Higgs mass is much less important than of the stop sector, $A_\tau$ is not severely constrained by the Higgs mass measurements. However, large $|A_\tau|$ destabilises also the ew potential as $A_t$ does. We checked the MSSM parameter space within the following ranges 
\begin{eqnarray}
&M_1 \in [0.1,1]~\text{TeV},\ M_2\in [0.15,1]~\text{TeV},\ M_3 \in [1.5,3]~\text{TeV}& \nonumber \\
&\mu \in [-2,2]~\text{TeV},\ M_A \in [0.2,2]~\text{TeV},\ \tan\beta \in [5,50]& \nonumber \\
&m_{\tilde{t}_L} \in [1,3]~\text{TeV}, m_{\tilde{t}_R} \in [0.2,3]~\text{TeV}& \nonumber \\
&m_{\tilde{\tau}_L} \in [0.1,1]~\text{TeV}, m_{\tilde{\tau}_R} \in [0.1,1]~\text{TeV}& \nonumber \\
&A_t, A_\tau \in [-10,10]~\text{TeV}, A_b=0&
\end{eqnarray}
We always found that the other constraints are more severe than the ones from perturbative unitarity. As example we show in Fig.~\ref{fig:MSSM_stau} the results for ($A_\tau$,$M_A$) plane. The other parameters are chosen as
\begin{eqnarray}
&M_1=0.6~\text{TeV}, M_2=0.33~\text{TeV}, M_3=2~\text{TeV}&\nonumber \\
&\mu=0.5~\text{TeV}, \tan\beta=50, A_t=3~\text{TeV}& \nonumber \\
&m_{\tilde{\tau}_L},m_{\tilde{\tau}_R} = 0.5~\text{TeV},\  m_{\tilde{t}_L},m_{\tilde{t}_R} = 2.0~\text{TeV}&
\end{eqnarray}
The entire parameter region with $|a^{\rm max}_0| > 0.1$ is already in conflict with vacuum stability. For this finding, it is not even necessary to use the numerical check of the one-loop effective potential with {\tt Vevacious}, but already the thumb rule \cite{AlvarezGaume:1983gj}
\begin{align}
\label{eq:thumb_stau}
A_\tau^2 >&  3 (m_{\tilde{\tau}_L}^2+ m_{\tilde{\tau}_R}^2+m_{H_d}^2 + \mu^2) 
\end{align}
rules out the point with $|a_0^{\rm max}|$ close to 0.5.\\

All in all, the checks for perturbative unitarity seem not the necessary in the MSSM once other constraints are included.

\subsection{NMSSM}
We turn now to the NMSSM, i.e. the MSSM extended by a gauge singlet superfield. We consider the version with a $\mathbb{Z}_3$ to forbid all dimensionful parameters in the superpotential. The superpotential reads
\begin{equation}
W_{\rm NMSSM} = \lambda \hat H_d \hat H_u \hat S + \frac{1}{3}\kappa \hat S^3 + W_Y \,,
\end{equation}
with the standard Yukawa interactions $W_Y$ as in the MSSM. The additional soft-terms in comparison to the MSSM are
\begin{equation}
- \mathcal L_{\rm soft} \supset \left(A_\lambda \lambda H_d H_u S + \frac13  A_\kappa \kappa S^3 + \text{h.c.} \right) + m_s^2 |S|^2\,,
\end{equation}
where we have used the common parametrisation for the trilinear soft terms
\begin{equation}
 T_\lambda = A_\lambda \lambda \,,\hspace{1cm} T_\kappa = A_\kappa \kappa\,.
\end{equation}
After electroweak symmetry breaking, the scalar singlet $S$ obtains a VEV $v_S$ which generates an effective Higgsino mass term $\mu_{\rm eff}$ 
\begin{equation}
\mu_{\rm eff} = \frac{1}{\sqrt{2}} \lambda v_S \,.
\end{equation}

\begin{figure}
\includegraphics[width=\linewidth]{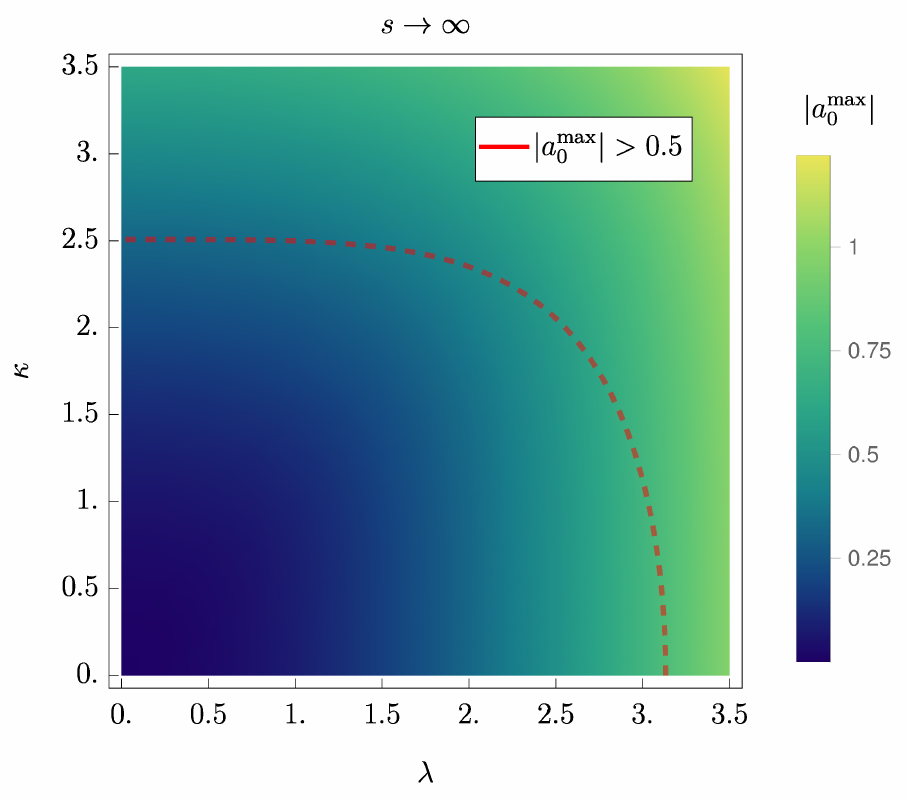} 
\caption{The value of $|a_0^{\rm max}|$ in the ($\lambda$,$\kappa$) plane in the large $s$ approximation. }
\label{fig:NMSSM_largeS}
\end{figure}

\begin{figure}[tb]
\includegraphics[width=\linewidth]{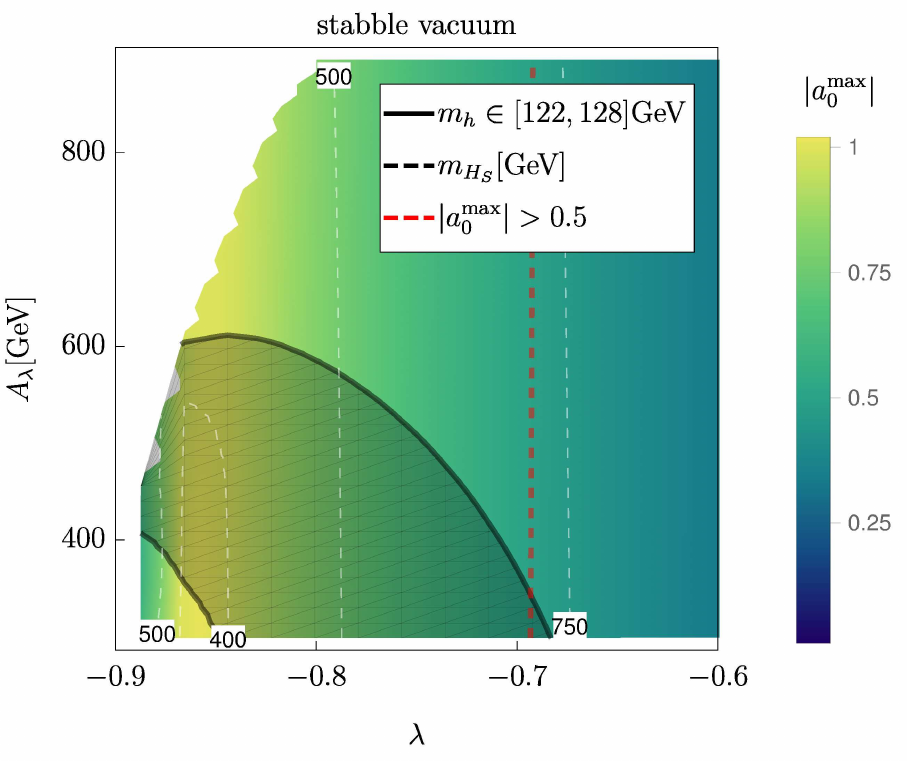} 
\caption{The values of $|a^{\rm max}_0|$ in the ($\lambda$,$A_\lambda$) plane. The red line indicates $|a^{\rm max}_0|>\frac12$. The grey shaded region shows the region with $m_h =125\pm 3$~GeV, while the white dashed line gives the mass of the CP even singlet. The entire plane has a stable ew vacuum. }
\label{fig:NMSSM1}
\end{figure}

\begin{figure}[tb]
\includegraphics[width=\linewidth]{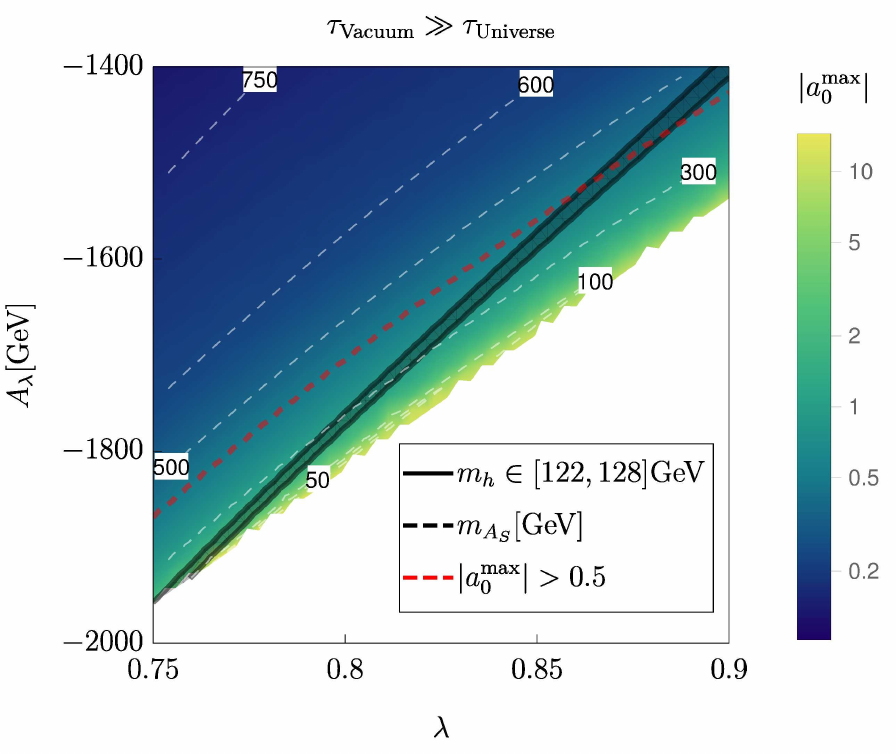} 
\caption{Similar to Fig.~\ref{fig:NMSSM1}, but this time showing the mass of the CP odd singlet by the white dashed line. The ew vacuum is metastable in the entire plane, but the life-time is many orders of magnitude bigger than the age of the universe. }
\label{fig:NMSSM2}
\end{figure}

\begin{figure}[tb]
\includegraphics[width=\linewidth]{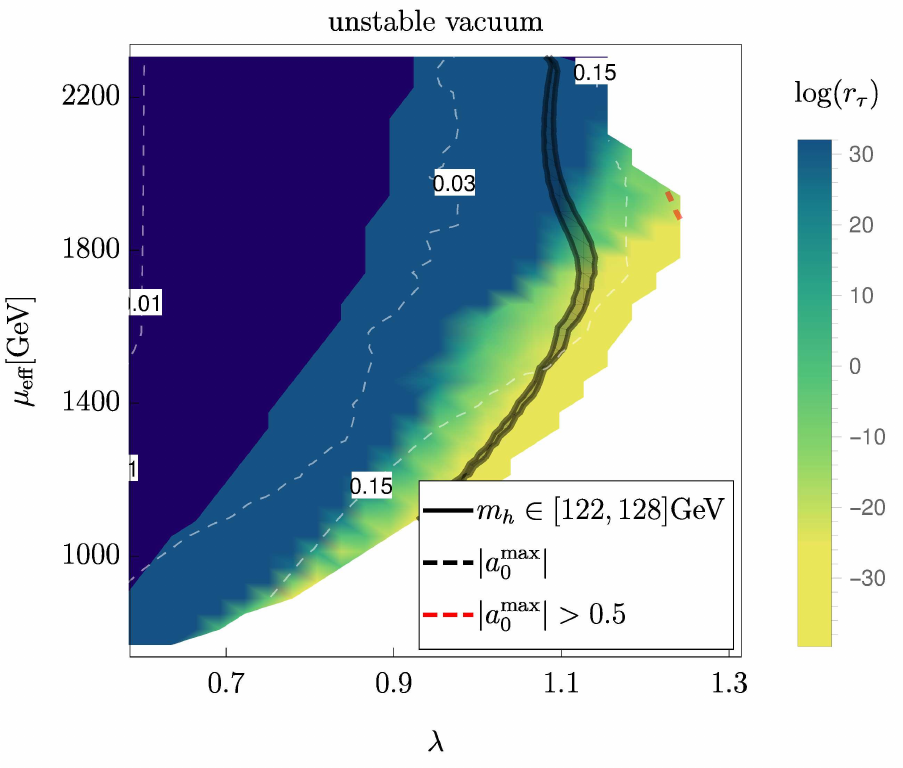} 
\caption{The life-time of the ew vacuum normalised to the age of the universe in the ($\lambda$,$\mu_{\rm eff}$)-plane. The white dashed lines show the values of constant $|a_0^{\rm max}|$, and the red line indicates  $|a_0^{\rm max}|>\frac12$. Again, the dark shaded region is preferred by the Higgs mass measurements. }
\label{fig:NMSSM3}
\end{figure}

In contrast to the MSSM\footnote{All point interactions in the MSSM are fixed by gauge and Yukawa couplings which are too small to cause problems with perturbative unitarity 
for reasonable values of $\tan\beta$.}, there are already limits from perturbative unitarity in the NMSSM even in the large $s$ limit in which only quartic couplings 
contribute. These limits are given by
\begin{eqnarray}
&8 \pi > \text{max}\Big\{\left| \lambda \right| ^2,2 | \sqrt{\kappa ^2+\lambda ^2} \sqrt{\left(\kappa ^*\right)^2+\left(\lambda ^*\right)^2}| , &  \nonumber \\
& \frac{1}{2} | 4 |\kappa|^2+|\lambda|^2 \pm \sqrt{16 |\kappa|^4-8 |\kappa|^2  |\lambda|^2 +17 |\lambda|^4}| , & \nonumber \\
&| 2 \text{Re}(\lambda  \kappa^*)\pm\sqrt{ 2\text{Re}(\lambda ^2 \kappa^{*,2})-2 |\kappa|^2  |\lambda|^2+|\lambda|^4}| \Big\}  &
\end{eqnarray}
and can be summarised in the plot shown in Fig.~\ref{fig:NMSSM_largeS}.
Thus, only if $|\lambda|$ and/or $|\kappa|$ have values well above 2, these constraints come into play. Such extreme parameter regions are rarely considered in phenomenological studies, i.e. 
these constraints are of a very limited, practical relevance. However, once the large $s$ limit is given up, one can find much stronger constraints because of new contributions involving $A_\lambda$, $A_\kappa$ 
as well as effective trilinear couplings proportional to $\mu_{\rm eff}$. 
We use the parameters as in eq.~(\ref{eq:default_values}) as far as applicable in the NMSSM and set in addition
\begin{eqnarray}
&\kappa = -2,\ A_\kappa = -2162~\text{GeV},\ \mu_{\rm eff} = 249~\text{GeV}& \nonumber \\
&\tan\beta=5.27,\  A_t=-1628~\text{GeV}&
\end{eqnarray}
When scanning over $\lambda \in [-0.9,0.6]$ as well $A_\lambda \in [300,900]$~GeV, we find the behaviour as shown in Fig.~\ref{fig:NMSSM1}: the values of $|a_0|$ are in general big and in particular the region with the desired Higgs mass is in conflict with the condition of perturbative unitarity. One can also observe that the value of $a_0^{\rm max}$ scales with the mass of the CP even singlet, i.e. the large amplitudes are caused by light singlet propagators. It is worth to stress that in the entire plane depicted in Fig.~\ref{fig:NMSSM1} the ew-vacuum is stable, i.e. without checking perturbative unitarity one would assume that the full parameter region with correct Higgs mass is viable.  
A similar situation is depicted in Fig.~\ref{fig:NMSSM2} where again $\lambda \in [0.75,0.9]$ and $A_\lambda \in [-2.0,-1.4]$~TeV has been varied while the other parameters where chosen as 
\begin{eqnarray}
&\kappa = 0.65,\ A_\kappa = -106~\text{GeV},\ \mu_{\rm eff} = 2823~\text{GeV}& \nonumber \\
&\tan\beta=1.13,\ A_t=1448~\text{GeV}&
\end{eqnarray}
This time, the ew vacuum is not stable, but the life-time is many orders of magnitude longer than the age of the universe. Thus, one might again assume that the shown parameter region is fine with respect to theoretical, and also experimental, constraints. However, the presence of a very light pseudo-scalar $A_S$ causes large scattering rates $hh \to hh$. Therefore, the interesting regions with $m_h \simeq 125$~GeV violates perturbative unitarity.  \\
Even if no CCB minima are present, one can also find the opposite situation in the NMSSM: parameter points might be in agreement with perturbative unitarity but suffer from an unstable vacuum. An example for this is given in Fig.~\ref{fig:NMSSM3} where the following parameters are used:
\begin{eqnarray}
&\kappa = 0.97,\ A_\lambda = -1064~\text{GeV},\ A_\kappa = -2454~\text{GeV}& \nonumber \\
&\tan\beta=1.26, \ A_t=1329~\text{GeV}&
\end{eqnarray}
In this plane only a tiny region is affected by the unitarity constraints, while the vacuum stability checks rule out a large fraction of the points in agreement with the correct mass for the SM-like Higgs. \\

\begin{figure}
\includegraphics[width=\linewidth]{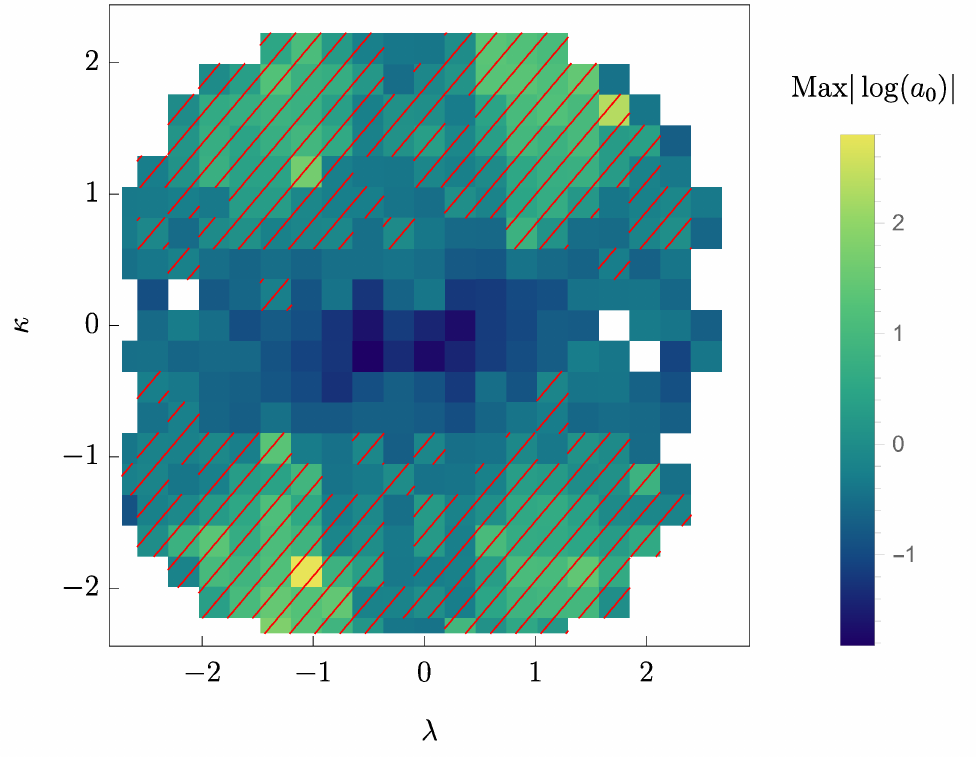}  \\
\includegraphics[width=\linewidth]{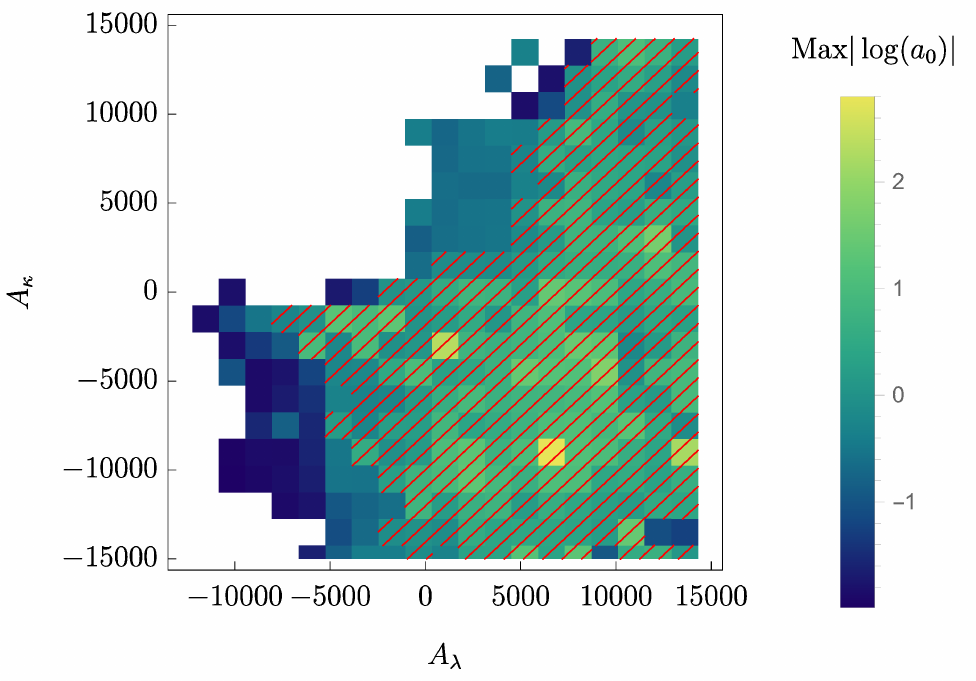}  
\caption{The maximal value of $|a_0^{\rm max}|$ per bin in for a scan in the parameter ranges given in eq.~(\ref{eq:NMSSM_ranges}). The red hatching shows where parameter points with $|a_0^{\rm max}|>\frac12$ were found, i.e. where perturbative unitarity could be violated.}
\label{fig:NMSSM_global}
\end{figure}

We have discussed so far selected planes where either vacuum stability or perturbative unitarity are the dominant constraint. Since the constraints from perturbative unitarity in the NMSSM rarely discussed up to now in literature, we want to give a brief impression of the overall situation. For this purpose, we summarise in Fig.~\ref{fig:NMSSM_global} the results of a parameter sample of 5 mio points in the following ranges
\begin{eqnarray}
& \lambda,\kappa \in [-3,3], \ A_\lambda,A_\kappa \in [-15,15]~\text{TeV} & \nonumber \\
& \mu_{\rm eff} \in [0.1,3]~\text{TeV}, \ \tan\beta \in [1.05,10.] & \nonumber \\
& M_1 \in [0.1,1.5]~\text{TeV}, \ M_2 \in [0.15,1.5]~\text{TeV}& \nonumber \\
& M_3 \in [1.5,3.0]~\text{TeV}& \nonumber \\
& m_{\tilde{t}_L},m_{\tilde{t}_R} \in [1,3]~\text{TeV},\ A_t \in [-2.5,2.5]~\text{TeV} & 
\label{eq:NMSSM_ranges}
\end{eqnarray}
The plots show the maximal value of $|a^{\rm max}_0|$ which we found in each bin. 
One can see that there are parameter combinations which are safe, i.e. perturbative unitarity is never violated. This is for instance the case if $|\kappa|$ is small.  
However, for the large majority of points there is not such a clear condition and a proper check of perturbative unitarity is necessary.

\section{Conclusion}
\label{sec:conclusion}
We have summarised in this letter the situation concerning the impact of perturbative unitarity and vacuum stability on the MSSM and NMSSM parameter spaces. We showed that the constraints from vacuum stability are important in the MSSM  because they can rule out phenomenological interesting parameter regions. In contrast, perturbative unitarity constraints in the MSSM just come into play once other constraints like vacuum stability or the Higgs mass are already at work. Thus, performing checks for perturbative unitarity in phenomenological studies seems not to be necessary as long as the other constraints are included. \\
The situation is completely different in the NMSSM. We have shown at a few examples that, depending on the considered parameter region, either checks for vacuum stability or perturbative unitarity are more important. While tests of the vacuum stability are already taken into account in NMSSM studies at least to some extent, perturbative unitarity is usually ignored. Therefore, we hope that the results in this letter demonstrate also the need for careful checks of perturbative unitarity in the NMSSM -- and most likely in many other, non-minimal SUSY models.

\section*{Acknowledgements}
I thank Mark Goodsell for many helpful discussions and a close collaboration in related projects.
This work is supported by ERC Recognition Award ERC-RA-0008 of the Helmholtz Association. 

%
%
%
%
%
%

\bibliography{lit}

\end{document}